# Force-Driven Growth of Intercellular Junctions[†]

**Mohammad Tehrani, Alireza S. Sarvestani**

*Department of Mechanical Engineering, Ohio University, Athens, OH 45701*

Submitted to

JOURNAL OF THEORETICAL BIOLOGY

**Corresponding Author:**
Alireza S. Sarvestani
Department of Mechanical Engineering
248 Stocker Center
Ohio University
Athens, OH 45701
Tel: (740) 593-9558
Fax: (740) 593-0476
sarvesta@ohio.edu

---

[†] Dedicated to Professor Mohsen Shahinpoor on occasion of his 74[th] birthday.

**Abstract**

Mechanical force regulates the formation and growth of cell-cell junctions. Cadherin is a prominent homotypic cell adhesion molecule that plays a crucial role in establishment of intercellular adhesion. It is known that the transmitted force through the cadherin-mediated junctions directly correlates with the growth and enlargement of the junctions. In this paper, we propose a physical model for the structural evolution of cell-cell junctions subjected to pulling tractions, using the Bell-Dembo-Bongard thermodynamic model. Cadherins have multiple adhesive states and may establish slip or catch bonds depending on the $Ca^{2+}$ concentration. We conducted a comparative study between the force-dependent behavior of clusters of slip and catch bonds. The results show that the clusters of catch bonds feature some hallmarks of cell mechanotransduction in response to the pulling traction. This is a passive thermodynamic response and is entirely controlled by the effect of mechanical work of the pulling force on the free energy landscape of the junction.

**Introduction**

Many important biological and pathological processes depend on the interactions between cells. Differentiation, embryogenesis, tumor invasion, and metastasis are often controlled by intercellular interactions (A. Pierres 2000; Derycke and Bracke 2004). Cell-cell adhesion is especially important in maintaining the tissue integrity and resisting the mechanical challenges experienced by epithelial and endothelial cells (Harris and Tepass 2010). The cell-cell adhesion is often established by specific homotypic interactions of dedicated transmembrane adhesion molecules. The major types of specialized intercellular junctions are tight junctions, gap junctions, and adherens junctions (AJs), mediated by occludins, connexins, and cadherins, respectively (Dejana 2004; Vasioukhin and Fuchs 2001).

The specific binding between cadherins is one of the key steps in establishment of cell-cell contacts (Baum and Georgiou 2011). Cadherin consists of an extracellular adhesive domain and a cytoplasmic domain, which is linked to the actin cytoskeleton by a battery of cytoplasmic signaling and scaffolding proteins, such as p120 phosphoprotein, β-catenin, α-catenin, and vinculin (Nagafuchi 2001; Yap et al. 1997). Linked to the actomyosin contractile machinery, the AJs provide an intercellular mechanical linkage to transmit the cytoskeletal contraction between the cells (Hoelzle and Svitkina 2012). Transmission of endogenic or exogenic forces through the AJs controls the growth and maturation of junctions and initiates the cascades of intracellular signals that are detrimental to the cellular phenotype (Chen et al. 2004). This is reminiscent of the well-known sensory function of cell-matrix focal adhesions (FAs) (Eyckmans et al. 2011). It is known that the FAs grow in response to the mechanical forces. A popular hypothesis holds that the FAs are able to perceive the mechanical stimulations and translate them into chemical signaling, a phenomenon often referred to as cell mechanotransduction (Vogel and Sheetz 2006).

Structure and functions of intercellular AJs share noticeable similarities with FAs. They both consist of dense clusters of transmembrane receptors that bind selectively to other cells or components of extracellular matrix. FAs and AJs are connected to the actin filaments through a myriad of scaffolding and plaque proteins that transmit the cytoskeletal traction to the adhesion site. These similarities motivated the idea that AJs may also act as mechanosensors (Schwartz and DeSimone 2008). Experiments have demonstrated that AJs grow in response to externally applied forces as well as internal actomyosin contraction. Liu et al. (Liu et al. 2010) reported a direct correlation between the magnitude of tugging forces transferred through endothelial cell junctions and the size of the junctions. They found that myosin II activity facilitates the bridge formation and accumulation of cadherins in AJs. le Duc et al. (le Duc et al. 2010) examined the response of AJs of epithelial cells to the externally applied forces. Beads coated with recombinant E-cadherin ectodomains were trapped in a magnetic field and adhered to the dorsal surfaces of cultured epithelial cells. The beads were then twisted by oscillating the magnetic field. They found that the adhesive contact between the bead and the cell stiffens in response to the repetitive twisting force. The strengthening of adhesion was particularly observed by further accumulation of vinculin, a protein involved in anchoring the actomyosin filaments to the adhesion junctions.

The strengthening mechanism of cell-cell junctions is critical for maintaining tissue integrity and preventing the cell-cell rupture in response to external forces. There is a growing interest in understanding the physical and molecular mechanisms that regulate the sensory pathways at the AJs. Several molecular mechanisms have been postulated with regards to the role of cytoplasmic proteins in regulating mechanotransduction at AJs. It is generally believed that the force transmission through binding partners of vinculin may alter their conformations and capacity to interact with vinculin. For example, it is proposed that the force-dependent recruitment of vinculin is induced by the unfolding of α-catenin and exposing its cryptic binding sites to vinculin (Yonemura et al. 2010).

From a physical viewpoint, it is important to understand how a single cadherin bond responds to increasing tension. A physically intuitive description of unbinding kinetics was first proposed by Bell (Bell 1978), which states that the average rate of unbinding, $k_{off}$, is exponentially dependent on the tensile force developed in the bond. This model characterizes the behavior of *slip bonds*; a family of biological bonds whose lifetime shortens with the application of tensile force. Alternatively, *catch bonds* represent the behavior of a group of biological bonds whose lifetime shows a biphasic dependence on applied mechanical tension and may become longer lived when subjected to mechanical tension (Zhu et al. 2005). The force-enhanced adhesion of catch bonds has been identified in several key cellular bonds (Marshall et al. 2003; Sarangapani et al. 2004; Truong and Danen 2009). Recently, the molecular mechanism of cadherin catch bond formation is demonstrated (Manibog et al. 2014; Rakshit et al. 2012). Accordingly, cadherins can form catch bonds in their x-dimer adhesion state. At this configuration, the application of mechanical force leads to formation of *de novo* hydrogen bonds that tighten the contact between the x-dimers.

It has been suggested that the establishment of catch bonds may regulate the mechanotransduction at cell adhesion sites (Schwarz 2013). Novikova and Storm (Novikova and Storm 2013) has presented an analytical assessment of the effect of catch bond formation between ligands and receptors on the behavior of FAs . Their results show that the clusters of catch bonds provide stronger adhesion to the extracellular matrix compared to the slip bonds. In the present work, we will extend their approach and conduct a comparative study between the force-dependent behavior of slip and catch bonds at an isolated adhesion site between two cells. To this end, we use a deterministic approach based on the classical Bell-Dembo-Bongard model for the adhesion at biological interfaces (Bell et al. 1984). Within this thermodynamic framework, membrane adhesion is mediated by the interplay between specific adhesions, generic interfacial repulsion, and the mixing entropy. We use this thermodynamic model to analyze the effect of pulling force on the population of bonds in a cluster and the size of the adhesion site.

**Adhesion Model**

Figure 1 schematically shows an adhesion site formed by a cluster of bonds between two similar cells with membrane areas of $A_1$ and $A_2$. The flexible cell membranes are conjugated via $N_b$ specific bonds spread out over an adhesion site with area of $A_b$. The initial number of laterally mobile receptors on the membranes is assumed to be $N_1$ and $N_2$, respectively. The short-range adhesive interaction between the homotypic receptors is opposed by the inter-membrane repulsion, induced by thermal fluctuation, dipole interactions, or steric repulsion. In the deterministic approach presented here, the repulsion is due to the steric repulsion imposed by a squeezed layer of glycocalyx covering the cell membranes. Membranes have a finite bending rigidity $\kappa$ and are subjected to membrane tension $\sigma$. The adhesion complex formed by the cluster of receptor-receptor bonds is assumed to survive long enough to mature into a junction after association with the scaffolding/adaptor proteins and actin microfilaments. The matured cluster is subjected to the pulling force $F$, due to motor activity of myosins.

The plaque proteins accumulate at the cytoplasmic side of the junction and are assumed to transmit the actomyosin generated force to the bonds. These bonds are modeled as simple harmonic springs with stiffness $k_b$ and their rest and deformed lengths are represented by $l_{b0}$ and $l_b$, respectively. The elastic membrane locally deforms around the established bonds to conform to the disjoining pressure induced by the inter-membrane repulsion. Here, we characterize the gap between the membrane profiles by $u(\boldsymbol{r})$ where $\boldsymbol{r}=(x,y)$ indicates the position vector on an arbitrary plane perpendicular to the bonds.

Various theoretical models are introduced to capture the catch-to-slip transition of biological bonds (Prezhdo and Pereverzev 2009; Sokurenko et al. 2008; Zhu et al. 2005). Here, we use the so-called *two-pathway* model to conceptualize the kinetics of bond dissociation (Pereverzev et al. 2005). This model has been fitted quantitatively to a number of experimental catch-bond data and provided useful analytical results. Within the framework of this model, bonds reside in a single stable state but may depart the energy well via two alternative pathways with different energy barriers. Figure 2 schematically shows the potential energy profile of a bond initially residing in state $x_e$, the equilibrium distance between the receptors. The slip and catch energy barriers correspond to the separation distance $x_s$ and $x_c$, respectively. In the absence of force, the slip pathway involves higher energy barrier compared to that of the catch state and thus, there is a higher probability to escape the potential well via the catch pathway. Development of force $f_b$ in the bond changes the energy landscape with different effects on bond configuration in the slip and catch states. In the slip state, force pulls the receptors away and decreases the energy barrier whereas in the catch state it pushes them in and increases the energy barrier. With increasing the force, taking the catch pathway becomes more costly and bonds are compelled to slip. The rate constant of transition from a bound to a free state is taken to be the superimposition of Kramer's type unbinding rates through slip and catch pathways. That is

$$k_{off}(f_b) = k_s \exp\left[\frac{f_b \delta_s}{k_B T}\right] + k_c \exp\left[\frac{-f_b \delta_c}{k_B T}\right]. \tag{1}$$

Here $\delta_s = |x_s - x_e|$, $\delta_c = |x_c - x_e|$ and $k_s$ and $k_c$ represent the force-independent rates of slip and catch unbinding, respectively. Pereverzev et al. (Pereverzev et al. 2005) evaluated the average lifetime of a bond during the catch to slip transition. Assuming that unbinding is a Poisson process, its probability density at time $t$ can be represented by $p(t) = k_{off} \exp[-k_{off} t]$. Thus, the mean lifetime of a bond subjected to force $f_b$ follows

$$\tau(f_b) = \int_0^\infty t\, p(t)\, dt = k_{off}^{-1}(f_b). \tag{2}$$

Eq. (2) has four parameters that can be obtained by fitting to the experimental data. Here for simplicity, we set the activation lengths $\delta_s$ and $\delta_c$ equal and represent them with a single length scale $\delta$ (Novikova and Storm 2013; Sun et al. 2012). Since bonds are taken to be linear springs, we further assume that $\delta = l_b - l_b^0 = f_b / k_b$. The binding kinetics of receptors can be represented as

$$\frac{d}{dt}\left(\frac{N_b}{A_b}\right) = -k_{off}\left(\frac{N_b}{A_b}\right) + k_{on}\left(\frac{N_1 - N_b}{A_1}\right)\left(\frac{N_2 - N_b}{A_2}\right). \tag{3}$$

The 2D binding rate $k_{on}$ is assumed to change with $\delta$ as (Sun et al. 2009)

$$k_{on} = k_{on}^0 \exp[-\delta^2 / \delta_{on}^2], \tag{4}$$

where $k_{on}^0$ is a binding coefficient and $\delta_{on}$ is a characteristic binding length.

The dominant mechanism for the growth of cell adhesion sites depends on the relative magnitude of the diffusion time of free receptors and the characteristic reaction time between receptors (Boulbitch et al. 2001). If the diffusion time far exceeds the time required for reaction, the association between the membranes is controlled by the flux of the receptor into the adhesion site. On the other extreme, if the reaction time is relatively long, the binding is reaction dominated and the diffusion plays a negligible role. The theory presented here is restricted to the case in which the diffusion of receptors and cytoplasmic plaque proteins happens on a timescale much faster than the characteristic time of receptor binding (Dembo et al. 1988; Olberding et al. 2010). This assumption, although restrictive, greatly simplifies the mathematical framework of our model. One important implication of this assumption is that it guaranties the balance of lateral osmotic pressure induced by the clustered bonds and the energy penalty to overcome the repulsion, as discussed later.

The total Gibbs free energy of the system, $G_{tot}$, is comprised from different contributions

$$G_{tot}(A_b, N_b, l_b) = U + G_b + G_g + G_m - W + G_{cfg}. \tag{5}$$

Here $U$ is the adhesion energy, $G_b$ is the stored energy in the stretched bonds, $G_m$ is the elastic energy stored in the flexible membrane, $G_g$ is the repulsive inter-membrane potential (e.g., induced by the compressed glycocalyx), and $W$ is the mechanical work done by the pulling force $F$. Additionally, $G_{cfg}$ represents the configurational contributions in free energy, defined as

$$G_{cfg} = -TS, \tag{6}$$

where $T$ is the temperature and $S$ shows the mixing entropy.

The total binding energy depends on $E_b$, the enthalpy gain associated with the binding of a pair of receptors, and the number of bonds; i.e., $U = U(N_b, E_b)$. The stored energy in stretched bonds can be presented by

$$G_b = \frac{1}{2}\sum_{i=1}^{N_b} k_b (u_i - l_{b0})^2, \tag{7}$$

where $u_i = u(r_i)$ shows the gap distance at $r_i$, the position vector of the $i^{th}$ bond. The inter-membrane repulsive interaction is taken to be a simple harmonic potential with strength $k_g$ placed at $u_g$ (Schmidt et al. 2012)

$$G_g = \int_{A_b} d^2r \left\{ \frac{1}{2} k_g \left( u_g - u(r) \right)^2 \right\}. \tag{8}$$

$G_m$ in Eq. (5) represents the elastic energy of the flexible membrane represented by the energy functional (Schmidt et al. 2012)

$$G_g = \int_{A_b} d^2r \left\{ \frac{\kappa}{2} \left( \nabla^2 u(r) \right)^2 + \frac{\sigma}{2} \left( \nabla u(r) \right)^2 \right\}. \tag{9}$$

The actomyosin force $F$ contributes to the extension of the elastic bonds and the mechanical work by

$$W = \int_{l_{b0}}^{l_b} F \, dl_b, \tag{10}$$

assuming that the pulling force is evenly distributed among bonds and all bonds are equally deformed (i.e., $u_i = l_b$). The free receptors may conjugate with other receptors and the resulting bonds adjust their position in the adhesion site in response to the pulling traction. The entropy cost of bond accumulation can be represented by

$$S = -k_B \left\{ (N_1 - N_b) \ln \left[ \frac{N_1 - N_b}{A_1} l_{b0}^2 \right] + (N_2 - N_b) \ln \left[ \frac{N_2 - N_b}{A_2} l_{b0}^2 \right] + N_b \ln \left[ \frac{N_b}{A_b} l_{b0}^2 \right] \right\}. \tag{11}$$

Note that here the area occupied by each receptor embedded in the membrane is approximated by $l_{b0}^2$.

Minimization of free energy with respect to $u$ leads to the following Euler-Lagrange equation for the gap distance between the membrane profiles in equilibrium

$$\kappa \nabla^4 u(\mathbf{r}) - \sigma \nabla^2 u(\mathbf{r}) + k_g \left( u(\mathbf{r}) - u_g \right) + \left( k_b \left( u(\mathbf{r}) - l_{b0} \right) - F/N_b \right) \delta(\mathbf{r} - \mathbf{r}_i) = 0, \quad (12)$$

where $\nabla^2$ is the Laplacian operator. Assuming that bonds are stretched equally, this equation can be solved subjected to the boundary conditions of

$$u(\mathbf{r}_i) = l_b, \text{ and } \nabla u(\mathbf{r}_i) = 0, \quad (13)$$

The solution to Eq. (12) is

$$u(\mathbf{r}) = u_g + C \sum_{i=1}^{N_b} \left( K_0 \left[ B_+ |\mathbf{r} - \mathbf{r}_i| \right] - K_0 \left[ B_- |\mathbf{r} - \mathbf{r}_i| \right] \right), \quad (14)$$

where $K_0$ is the modified Bessel function of the second kind,

$$B_\pm = \sqrt{\frac{\sigma \pm \sqrt{\sigma^2 - 4\kappa k_g}}{2\kappa}}, \quad (15)$$

and

$$C = \frac{u_g - l_b}{\sum_{i=1}^{N_b} \left( K_0 \left[ B_- |\mathbf{r}_i| \right] - K_0 \left[ B_+ |\mathbf{r}_i| \right] \right)}. \quad (16)$$

After partial integration of Eq. (12), we can derive the condition for mechanical equilibrium and find the tension developed at each stretched bond as

$$f_b = k_b \left( l_b - l_{b0} \right) = \frac{1}{N_b} \left( \int_{A_b} d^2 r \left\{ k_g \left( u_g - u \right) \right\} + F \right). \quad (17)$$

Further, the combination of repulsive potential and the stored elastic energy in the membrane can be simplified as (Appendix A)

$$G = G_g + G_m = \frac{\pi k_g N_b \left( u_g - l_b \right)^2 \left( \frac{1}{B_+^2} - \frac{1}{B_-^2} \right)}{\sum_{i=1}^{N_b} \left( K_0 \left[ B_+ |\mathbf{r}_i| \right] - K_0 \left[ B_- |\mathbf{r}_i| \right] \right)}. \quad (18)$$

In the limit of high density of accumulated bonds, the 2D bond distribution is approximated by a hexagonal-lattice. We made this assumption primarily to simplify the calculation of stored energy function of deformed membranes. However, the formation of high density and ordered distributions of cadherin bonds have been recently confirmed in AJs (Strale et al. 2015; Wu et al. 2015). Under this condition, the contribution of membrane elastic energy is screened and Eq. (18) can be estimated by the following simple expression (Appendix A)

$$G = \frac{1}{2} A_b k_g \left(u_g - l_b\right)^2, \text{ for } B_\pm \lambda << 1, \tag{19}$$

where $\lambda \approx \sqrt{A_b / N_b}$ is the average distance between the bonds. Henceforth, we shall use this approximation for our calculations (see Results and Discussions).

Inside the adhesion site, bonds are constantly formed and broken due to their finite lifetime and in response to the cytoskeletal traction. This is inherently a non-equilibrium process, as shown by Eq. (3). Additionally, the arrangement of established bonds at cell-cell junctions is dynamic and changes with time. The dynamics and mobility of cadherin bonds inside mature AJs are measured by photobleaching and photoactivation techniques (Daneshjou et al. 2015; Hoffman and Yap 2015; Yamada et al. 2005) and the results point to continuous rearrangement of cadherin linkages within AJs. The growth of adhesion sites is controlled by the difference between the lateral osmotic pressure induced by the bonds and the compressive pressure of inter-membrane repulsion. The generated lateral pressure can be expressed as (Appendix A)

$$\Pi = -\frac{1}{2} k_g (u_g - l_b)^2, \tag{20}$$

When the total free energy is minimized (i.e., $\partial G_{tot}/\partial A_b = 0$), the repulsive pressure induced by compressed glycocalyx balances the osmotic pressure of bonds

$$\frac{N_b k_B T}{A_b} = \frac{1}{2} k_g \left( u_g - l_b \right)^2. \tag{21}$$

When the system is out of equilibrium, $A_b$ and $N_b$ will spontaneously change until the state of minimum free energy is reached. If $N_b$ is kept constant, $\Pi$ acts as a self-assembly pressure to reduce $A_b$ by pushing bonds closer to each other. This is an example of repulsion-induced aggregation, first noted by Bruinsma et al. (Bruinsma et al. 1994). If $N_b$ decreases, the adhesion area will shrink, leading to membrane dewetting. An increase in $N_b$, however, gives rise to a larger adhesion area and growth of the cluster. In the present work, the diffusion of receptors and the transport of bonds are assumed to happen quickly enough so that the balance Eq. (21) is satisfied at any instant of time. Finally, it is safe to assume that the total of tensile forces developed in the stretched bonds balances the pulling force and the inter-membrane repulsion

$$F + A_b k_g \left( u_g - l_b \right) = N_b k_b \left( l_b - l_{b0} \right), \tag{22}$$

which is equivalent to $\partial G_{tot}/\partial l_b = 0$.

**Results and Discussions**

Table 1 shows the non-dimensional forms of the parameters. The values of the model parameters used in this study are listed in Table 2. The stiffness of the cadherin bonds has been studied using different experimental measurements and computational simulations (Baumgartner et al. 2000; Perret et al. 2004; Sivasankar et al. 2001; Sotomayor and Schulten 2008). The results greatly differ for different members of the cadherin family and strongly depend on the $Ca^{2+}$ concentration. Here, the bond stiffness is assumed to vary between 2-4 pN/nm, consistent with the reported values for E-cadherin using single molecular force spectroscopy (Bajpai et al. 2009). The lifetimes of cadherin bonds reported in the literature vary in a significantly wide range (Leckband and Sivasankar 2012). Here, we used Eq. (1) to present the unbinding rate of catch-slip bonds with $k_s = 0.15$ s$^{-1}$ and $k_c = 10$ s$^{-1}$. For slip bonds, the unbinding rate is estimated by

$$k_{off}(f_b) = k_s \exp\left[\frac{f_b \delta_s}{k_B T}\right]$$ with $k_s = 0.8$ s$^{-1}$. The resulting lifetimes are shown by Figure 3. These input parameters are chosen to yield lifetimes qualitatively similar to the experimental results of Manibog et al. (Manibog et al. 2014) and at the same time keep the computational time of our simulations relatively small.

At the first step, we find the equilibrium configuration of a junction subjected to a fixed pulling force. The equilibrium solution of state equations (3), (21) and (22) for a given pulling force leads to one stable local minimum for the free energy and one unstable saddle point. Figure 4 shows some representative examples of bond arrangement in the cluster and the gap profile between the membranes. For our calculations, we assumed that bonds are densely packed ($B_\pm \lambda << 1$) and used the asymptotic expression (19) to evaluate $G$, the summation of store elastic energy in the deformed membrane and the squeezed glycocalyx. Figure 5 shows the variation of $G$ with the bond length at a stable configuration. The results are compared with those obtained from the series expansion (18). Evidently, the series is rapidly convergent and can be safely replaced by the asymptotic solution with a negligible error.

The stable and unstable branches of the equilibrium solutions are shown in Figure 6 and 7, considering bonds with different rigidities. The pulling force tends to separate the membranes and will eventually lead to disassembly of the junction. The disassembly happens at a finite adhesion area between the membranes and is due to a saddle point bifurcation at a critical force, where the stable and unstable branches converge. We performed a comparative bifurcation analysis of the behavior of slip and catch bonds at the junctions, as shown by the figures. The gap distance between the membranes grows larger as the pulling force increases and stretches the bonds. The most striking difference between the behavior of slip and catch bonds appears by comparing the variation of $N_b$ with $F$. While the number of slip bonds is a decreasing function of the pulling force, the population of catch bonds grows with the force before junctional instability. As previously noted (Novikova and Storm 2013; Schwarz 2013), this is the characteristic behavior of catch bonds subjected to a shared force. Increasing the force makes the catch bonds longer-lived and allows the cluster to retain a larger number of bonds.

Another interesting observation is the strong and positive correlation between the pulling force and $A_b$, the size of the adhesion site, when dissociation takes place by the catch-slip transition. While slip transition leads to only a small increase in $A_b$, catch bonds appear to act as force transducer leading to the enlargement of the cluster. Qualitatively, this is similar to the results of Liu et al. (Liu et al. 2010) who demonstrated strong correlation between the tugging force and the size of AJs of endothelial cells. Force-induced enlargement of the adhesion area is more pronounced at lower $k_b$ values. The growth of adhesion sites with force predicted by our model is a purely thermodynamic response and is a result of the balance between the lateral osmotic pressure of accumulated bonds and the energy penalty to overcome the repulsion. The inter-membrane repulsion is alleviated by increasing the pulling force and the separation of membranes. This pressure drop is compensated by increasing the adhesion area and reduction of osmotic pressure. It should be noted that increasing the number of bonds does not results in a higher bond density at the junction. Instead, the areal densities of both slip and catch bonds slightly decay at higher pulling forces. The correlation between the force and $N_b$ (and $A_b$) depends on $k_c$ and becomes stronger as the unbinding via catch pathway is expedited (Figure 8). In obtaining the results presented here, we assumed that the activation length $\delta$ in the two-pathway model is force-dependent and equal with $\delta = l_b - l_b^0 = f_b/k_b$, as proposed by Sun et al. 2012. Alternatively, it can be argued that the length scale $\delta$ is a characteristic property of the bond and its value is independent from the pathway in the phase space. While this could be a valid argument, implementing a force-independent activation length for the bonds does not qualitatively change our results.

In the second part of our numerical study, we consider the kinetics of the force-dependent configuration of a junction. The pulling force is assumed to be a simple smooth function of time as (see the inset of Figure 9(a))

$$\bar{F}(\bar{t}) = \begin{cases} \bar{F}_{ini}, & \bar{t} \leq 10^5 \\ \bar{F}_{ini} + \dfrac{\bar{F}_{ult}}{2}\left(1 - \cos\left[\dfrac{\pi}{8\times 10^5}(\bar{t}-10^5)\right]\right), & 9\times 10^5 \geq \bar{t} \geq 10^5 \\ \bar{F}_{ini} + \bar{F}_{ult}, & \bar{t} \geq 9\times 10^5 \end{cases} \quad (23)$$

The equilibrium configuration of the junction, in absence of an external force, is taken as the initial condition for the kinetic Eq. (3). To ensure that the cluster's structure evolves along the stable branch of solutions, a small pulling force ($\bar{F}_{ini}=100$) is applied to the junction in a period of time during which the solutions fluctuates between the branches but eventually stabilizes. Thereafter, the pulling force is slowly and gradually increased until it reaches $\bar{F}_{ini} + \bar{F}_{ult}$ and stays constant. Figures 9 and 10 show the variation of state parameters with time for three different values of $\bar{F}_{ult}$. The highest value of $\bar{F}_{ult}$ is chosen to be equal to the critical force at the saddle point bifurcation. As expected, all results become unstable at this force.

The population of slip bonds decreases with time and reaches a constant value, following a trend opposite to that of the applied pulling force. When the ultimate force is comparable with the critical force at the saddle point, the cluster becomes unstable and number of bonds rapidly drops. Conversely, the number of the catch bonds raises with the force, concurrent with increasing their lifetime (Figure 11). Similarly, the adhesion area of the catch bond clusters strongly correlates with force and the cluster grows in size as the pulling force increases with time. The size of adhesion cluster made by slip bonds, however, follows a different trend. It slightly increases initially and then decays with stronger pulling forces.

A major assumption behind the presented theory is that the growth of adhesion cluster is reaction-dominated. The transport of free receptors is assumed to happen sufficiently fast so that the population of newly formed bonds is determined by Eq. (3) and is not restricted by the supply of free receptors by diffusion. This can be better understood by looking at the variation of $G_{cfg}$ with $N_b$, as shown by Figure 12. The changes in mixing entropy of clusters made by slip and catch bonds follow opposite trends as the junction evolves with time. The primary effect of pulling force is to contribute to mechanical work. In the case of slip bonds, this is followed by the reduction of surviving bonds, determined by Eq. (3), and a net enthalpy loss. The rising number of free bonds leads to an increase in mixing entropy. In the case of catch bonds, the mechanical works is accompanied by an enthalpy gain as more receptors contribute in binding process. Since the adhesion is reaction-controlled, the consumption of free receptors does not lead to a non-uniform distribution of free receptors near the edge of junction, as shown in diffusion-dominated cases (Shenoy and Freund 2005). Thus, the application of force downgrades the effect of mixing entropy before the emergence of saddle node bifurcation.

**Concluding Remarks**

In this paper we used a physical model to study the intercellular junctions subjected to an external pulling force. The study is relevant to the cluster of cadherin bonds at AJs subjected to a pulling force. The model is an extension of the classical Bell-Dembo-Bongard model and based on multiple simplifying assumptions. Probably the most tenuous assumption here is the lack of any sensory function for the myriad of the cytoskeletal plaque proteins. Instead, they are assumed to simply link the receptors to the cytoskeleton. The model further assumes that the diffusion of receptors happens sufficiently fast and the lateral osmotic pressure of accumulated bonds and the inter-membrane repulsion always remain in balance. The model is comprised from two elastic membranes with specific interactions, mediated by mobile receptors, and a simple repulsive potential. The localized pulling force at the junction modifies the free energy landscape of the junction as a thermodynamic system. We studied the equilibrium and kinetic configurations of the junction considering the slip or catch behavior for the clustered bonds. The clusters of catch bonds specifically showed some characteristics of cell mechanotransduction where the population of bonds and the adhesion area of junction grew spontaneously in response to the pulling traction.

Obviously, the proposed minimal model does not do the justice to the rich and complex repertoire of molecular structure at cellular junctions and does not capture many subtleties of the structure of AJs. For example, it does not include the contribution of the signaling proteins in the regulation of the cytoskeletal contraction nor does it represent the hierarchical structure of AJs. Cadherin organization in AJs is suggested to be hierarchical and vary over different length scales. AJs are formed by assembly of multiple micro-clusters, each of which contains smaller and more densely packed cluster of cadherin bonds (Yap et al. 2015). Our model represents the structural evolution of a single cadherin cluster and does not account for such hierarchy in structure. Direct comparison with experimental results such as those presented by Liu et al. (Liu et al. 2010) requires a more comprehensive model of AJ structure at different scales.

Cell mechanotransduction at cell-cell junctions is generally attributed to the putative sensory function of binding partner of vinculin such as α-catenin in the bundle of plaque proteins (Barry et al. 2014; Yonemura et al. 2010). Our physical model, on the other hand, proposes an entirely different scenario based on the catch-to-slip transition of cadherin bonds during dissociation. Our results show the pulling force transmitted at a cellular junction directly affects the accumulation of catch bonds formed by the homotypic receptors and leads to the enlargement of cluster. In this thermodynamic framework, the enlargement and apparent stiffening of adhesion sites are results of a thermodynamic process to achieve the state of minimum free energy, independent from any regulatory function of other molecular sensors as postulated in the other models. These thermodynamic contributions can be considered as a precursor for mechanotransduction at cell-cell junctions. At some point, they will be coupled with the biochemical machinery of the cell and become part of the signal transduction pathway.

These results encourage us to think of mechanotransduction as a general thermodynamic process, favored by establishment of catch bonds, and not necessarily specific to cadherin-mediated AJs of epithelial tissue cells. Different types of cell-cell junctions with entirely different molecular repertoire may feature the hallmarks of mechanotransduction that can be explained by the proposed model. For example, it is known that the structure of desmosomes in epidermal keratinocytes depends on the mechanical stimulations (Reichelt 2007; Russell et al. 2004; Takei et al. 1998; Yano et al. 2004). It is still unclear whether desmogleins or desmocollins can form catch bonds. But if catch bonds are present, our model suggests that the pulling traction exerted by the intermediate filaments could mediate the mechanotransduction and lead to spontaneous accumulation of desmogleins and desmocollins and enlargement of desmosomes.

## Acknowledgments

The authors are indebted to anonymous reviewers for their insightful comments.

# Appendix A

The summation of elastic energy of the membrane and the repulsive potential is

$$G = G_m + G_g = \int_A d^2r \left\{ \frac{\kappa}{2} \left(\nabla^2 u(r)\right)^2 + \frac{\sigma}{2} \left(\nabla u(r)\right)^2 \right\} + \frac{k_g}{2} \int_A d^2r \left\{ \left(u(r) - u_g\right)^2 \right\}, \quad (A1)$$

Using partial integration along with the equilibrium Eq. (17), we find

$$G = \frac{k_g}{2}(u_g - l_b) \int_A d^2r \left\{ u_g - u(r) \right\}. \quad (A2)$$

Substitution of the membrane profiles, given by Eq. (14), into (A2) yields

$$G = (l_{b0} - u_g) C k_g \pi \int_0^\infty dr \left\{ \sum_{i=1}^{N_b} \left( r K_0 [B_+ |r - r_i|] - r K_0 [B_- |r - r_i|] \right) \right\}. \quad (A3)$$

Using $\int_0^\infty dr \left\{ \sum_{i=1}^{N_b} r K_0 [B_\pm |r - r_i|] \right\} = \frac{N_b}{(B_\pm)^2}$, we obtain

$$G = \frac{N_b \pi (u_g - l_b)^2 k_g \left( \frac{1}{B_+^2} - \frac{1}{B_-^2} \right)}{\sum_{i=1}^{N_b} \left( K_0 [B_+ |r_i|] - K_0 [B_- |r_i|] \right)}. \quad (A4)$$

In the limit of $|B_\pm \lambda| \ll 1$, corresponding to the dense clustering, the denominator of Eq. (A4) can be approximated by an integral as

$$\sum_{i=1}^{N_b} K_0(B_\pm r_i) \approx 2\pi \int_0^\infty dz\, z K_0(z \lambda B_\pm) = \frac{2\pi}{(\lambda B_\pm)^2}. \quad (A5)$$

Substitution of (A5) into (A4) yields

$$G = \frac{1}{2} A_b k_g (u_g - l_b)^2. \quad (A6)$$

This energy provides a lateral self-assembly pressure of $\Pi = -\frac{\partial G}{\partial A_b} = -\frac{1}{2} k_g (u_g - l_b)^2$.

# Table Captions

**Table 1.**    List of non-dimensional parameters.

**Table 2.**    Values of model parameters.

# Figure Captions

**Figure 1.** (a) Schematic presentation of a cluster of bonds between two cells, subjected to pulling force $F$. (b) The flexible membranes are subjected to membrane tension and inter-membrane repulsive potential (e.g., induced by glycocalyx). Adhesion is mediated by the specific binding of receptors. The bonds are linear springs with length $l_b$ and stiffness $k_b$. The adhesion cluster is assumed to be circular in shape with the area of $A_b$ (not shown in the figure).

**Figure 2.** The profile of potential energy in the two-pathway model (Prezhdo and Pereverzev 2009). In the absence of force (solid line), bond resides in the equilibrium state $x_e$. It may escape the potential well via two pathways with different energy barriers at $x_c$ and $x_s$, corresponding to the catch and slip states, respectively. Development of mechanical force, $f_b$, changes the energy landscape and the energy barriers (dashed line).

**Figure 3.** The variation of lifetime of a single slip and catch bond with force ($\bar{k}_b = 100$).

**Figure 4.** The arrangement of bonds and the gap distance between the membranes at $\bar{F} = 58000$ for junctions formed by slip and catch bonds.

**Figure 5.** Examining the convergence of series expansion (18), plotted versus the bond length, at a stable configuration and in the absence of a pulling force. Blue line shows the results obtained by the asymptotic relation (19) in comparison with the series expansion (18) represented by dashed red lines. The number of terms in the series expansion is gradually increased considering more layers of bonds, as shown in the inset.

**Figure 6.** Variation of (a) $\bar{l}_b$, (b) $\bar{A}_b$, (c) $N_b$, and (d) $\bar{n}_b$ with pulling force at a junction formed by a cluster of slip bonds. Stable and unstable branches are shown by solid and dashed lines, respectively.

**Figure 7.** Variation of (a) $\bar{l}_b$, (b) $\bar{A}_b$, (c) $N_b$, and (d) $\bar{n}_b$ with pulling force at a junction formed by a cluster of catch bonds. Stable and unstable branches are shown by solid and dashed lines, respectively.

**Figure 8.** The effect of unbinding rate (via catch pathway) on the variation of (a) $\bar{A}_b$ and (b) $N_b$. Stable and unstable branches are shown by solid and dashed lines, respectively.

**Figure 9.** The variation of (a) $\bar{l}_b$, (b) $\bar{A}_b$, and (c) $N_b$ of a stable cluster of slip bonds with time ($\bar{k}_b = 100$). The inset on panel (a) shows the variation of the pulling force with time, according to Eq. (23), with three different values of $\bar{F}_{ult}$, labeled as (1), (2), and (3). The largest value of $\bar{F}_{ult}$ is equal to the critical force at the saddle point bifurcation.

**Figure 10.** The variation of (a) $\bar{l}_b$, (b) $\bar{A}_b$, and (c) $N_b$ of a cluster of catch bonds with time. See the caption of Figure 7 for more explanation.

**Figure 11.** Variation of the lifetime of a single slip or catch bond with time ($\bar{k}_b = 100$). The pulling force follows Eq. (23) and $\bar{F}_{ult}$ is set equal to the critical force at the saddle point bifurcation.

**Figure 12.** Variation of configurational components of free energy with the number of slip and catch bonds in a junction ($\bar{k}_b = 100$). The pulling force follows Eq. (23) and $\bar{F}_{ult}$ is set equal to the critical force at the saddle point bifurcation.

# Table 1

| Symbol | Expression | Definition |
|---|---|---|
| $\bar{l}_b$ | $l_b/l_{b0}$ | length of bond |
| $\bar{u}$ | $u/l_{b0}$ | gap profile between membranes |
| $\bar{x}, \bar{y}$ | $x/l_{b0}, y/l_{b0}$ | position coordinates |
| $\bar{A}_b$ | $A_b/l_{b0}^2$ | surface area of a junction |
| $\bar{n}_b$ | $N_b/\bar{A}_b$ | density of bonds |
| $\bar{k}_b$ | $k_b l_{b0}^2/k_B T$ | stiffness of each bond |
| $\bar{t}$ | $k_{on}^0 t/l_{b0}^2$ | time |
| $\bar{\tau}$ | $k_{on}^0/k_{off} l_{b0}^2$ | lifetime of each bond |
| $\bar{k}_c$ | $k_c l_{b0}^2/k_{on}$ | unbinding rate via catch pathway |
| $\bar{F}$ | $F l_{b0}/k_B T$ | total pulling force |
| $\bar{f}_b$ | $f_b l_{b0}/k_B T$ | force at each bond |
| $\bar{G}$ | $G/k_B T$ | stored energy in membrane and glycocalyx |
| $\bar{G}_{cfg}$ | $G_{cfg}/k_B T$ | configurational component of free energy |

# Table 2

| Parameter | Value | Source |
|-----------|-------|--------|
| $N_{1,2}$ | $10^4$ | (Bell et al. 1984) |
| $l_{b0}$ | $10\,nm$ | (Bell et al. 1984) |
| $u_g$ | $30\,nm$ | (Zuckerman and Bruinsma 1998) |
| $\delta_{on}$ | $10\,nm$ | (Sun et al. 2012) |
| $A_{1,2}$ | $10^2\,\mu m^2$ | Assumed |
| $k_g$ | $10^{-9}\,pN/nm^3$ | (Wiesinger et al. 2013)• |
| $k_b$ | $2-4\,pN/nm$ | (Bajpai et al. 2009) |
| $\sigma$ | $0.2\,pN/nm$ | (Lieber et al. 2013) |
| $\kappa$ | $50\,k_B T$ | (Zuckerman and Bruinsma 1998) |
| $k_{on}^0$ | $\sim 1\,\mu m^2/s$ | (Sun et al. 2012) |

• Their results are used to estimate the value of $k_g$.

Figure 1

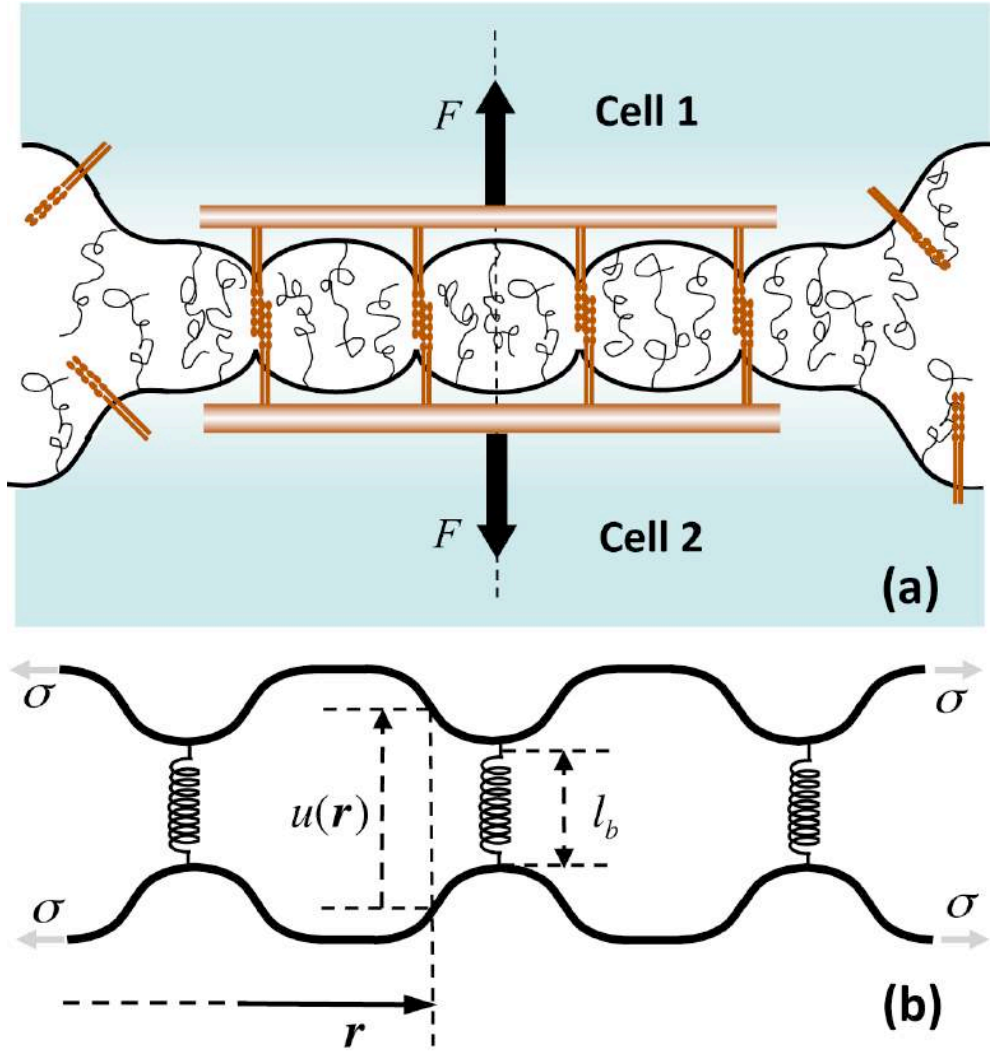

Figure 2

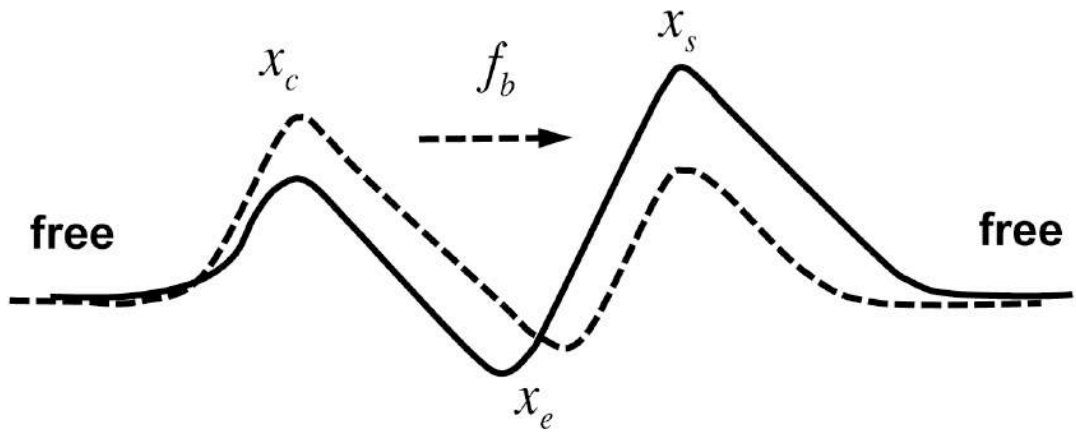

Figure 3

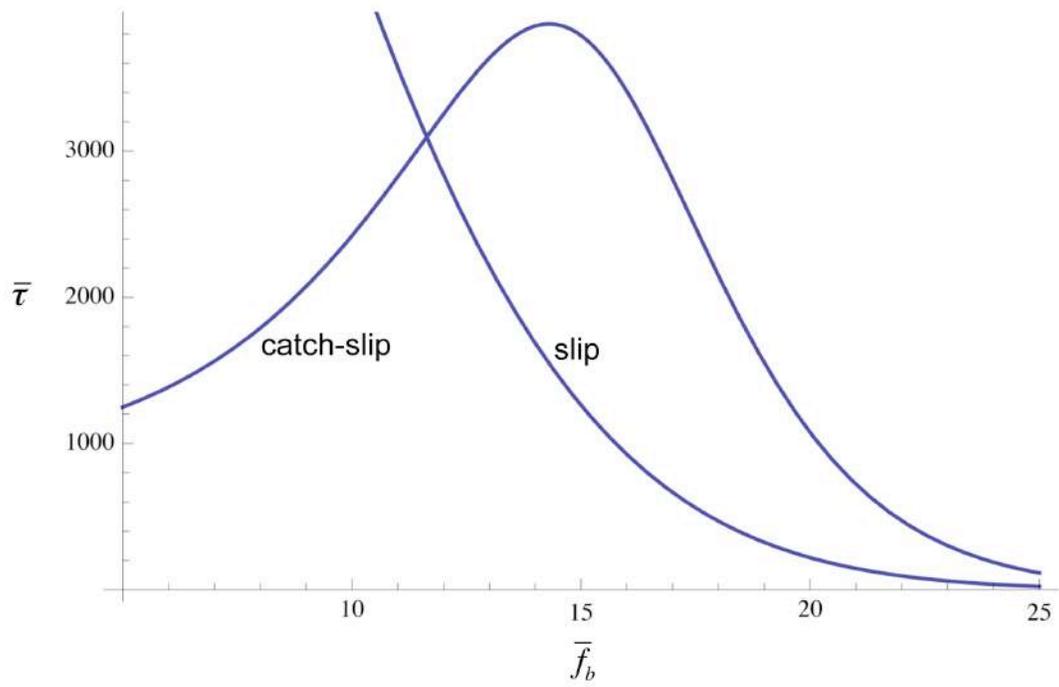

Figure 4

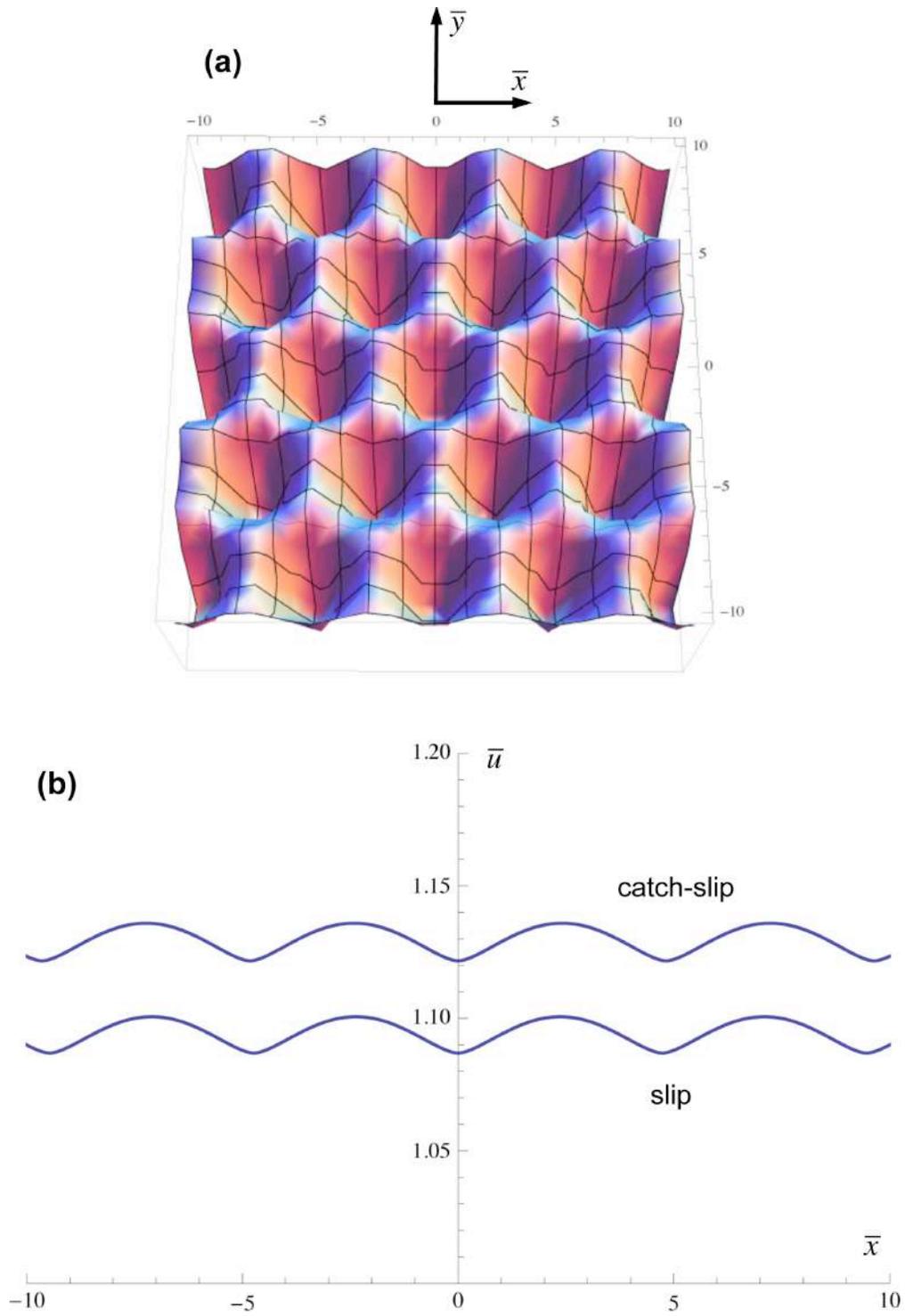

Figure 5

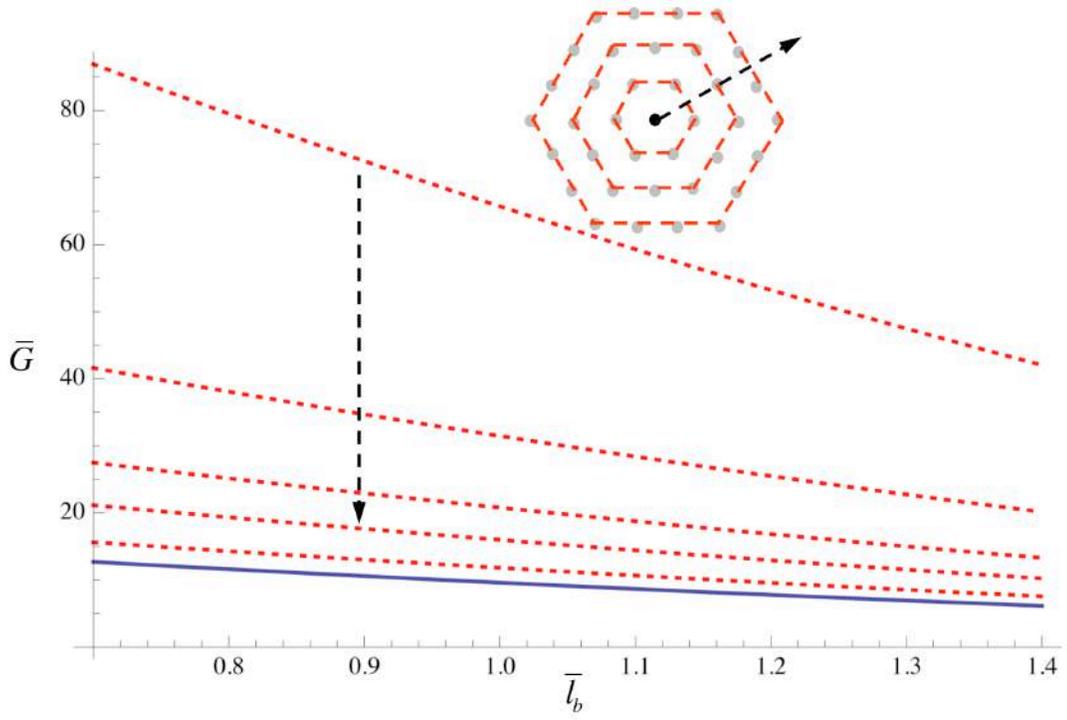

Figure 6

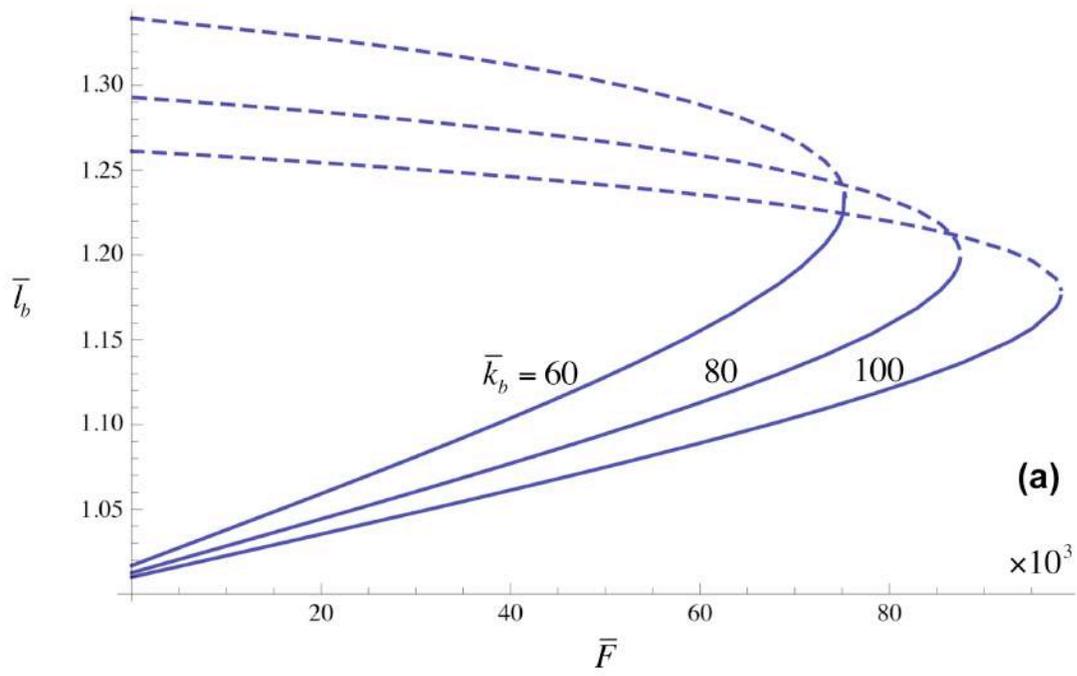

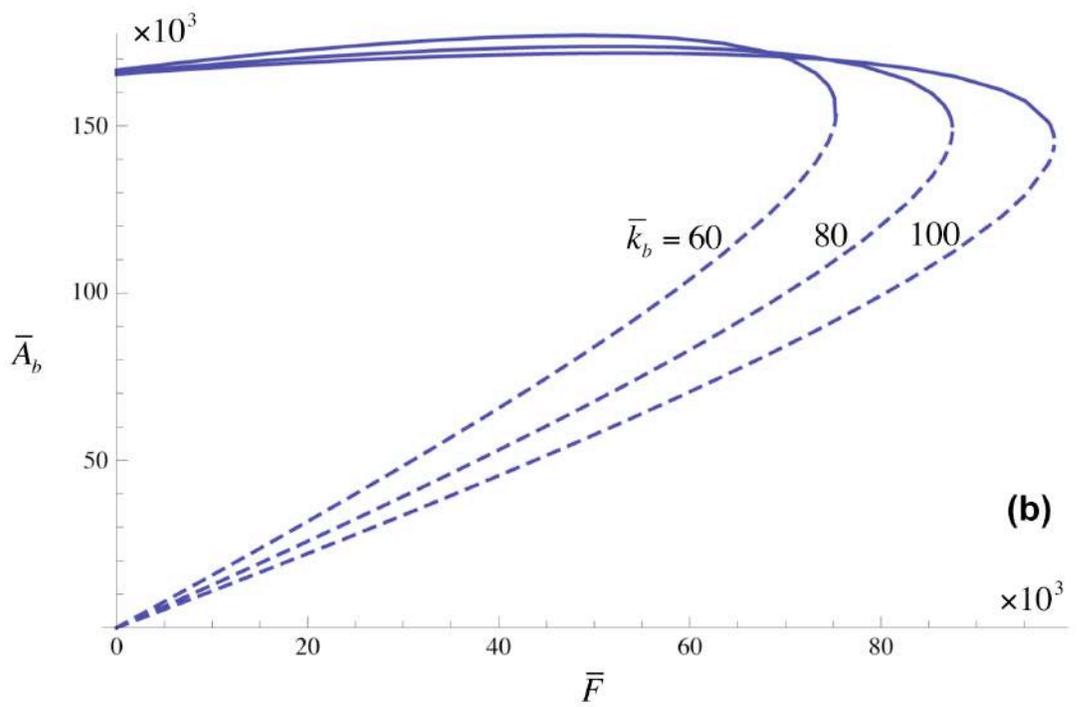

Figure 6

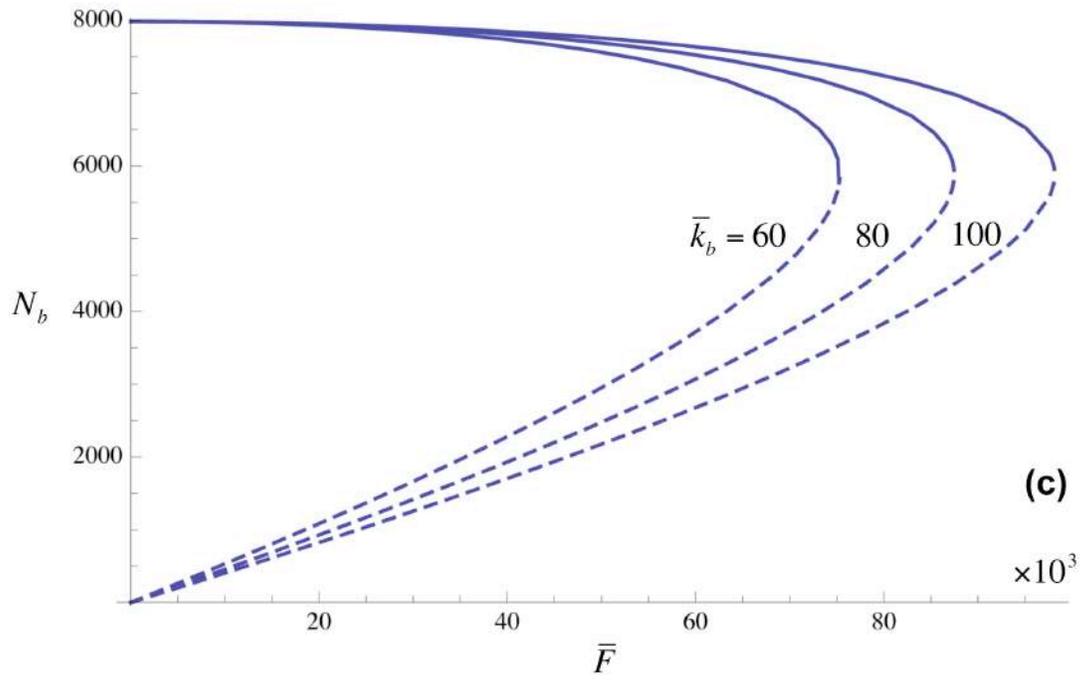

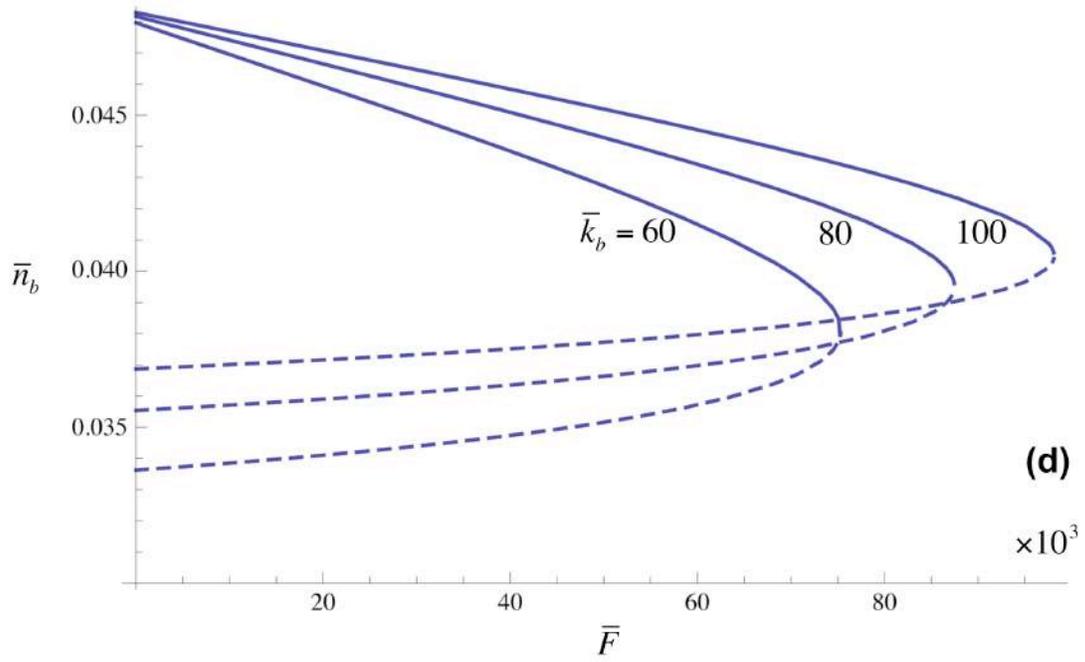

Figure 7

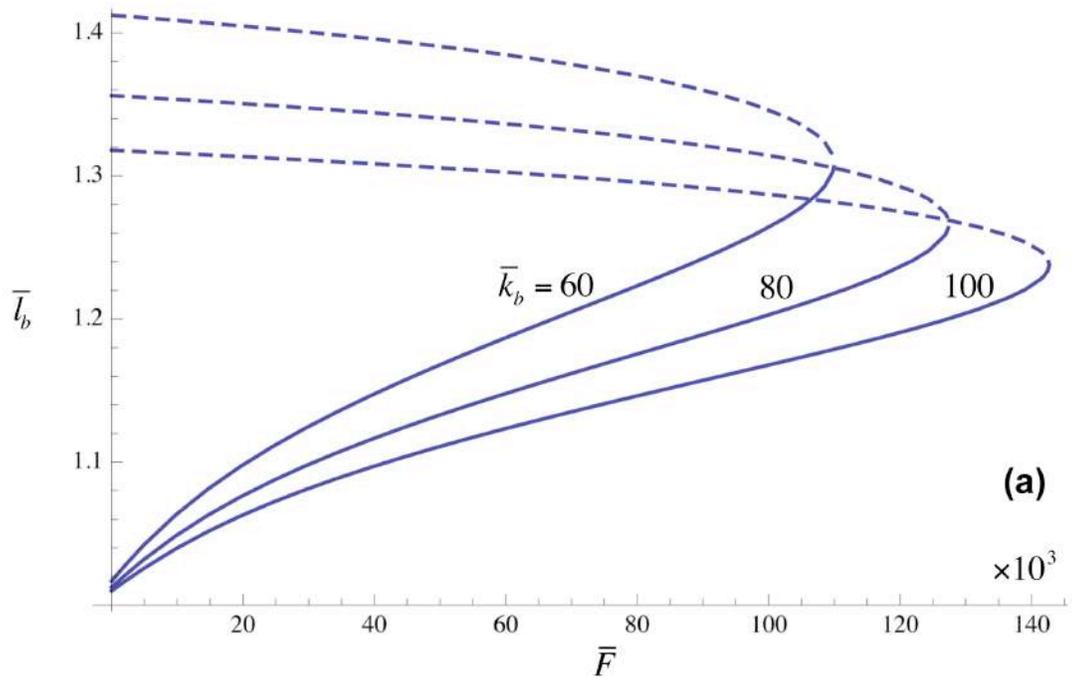

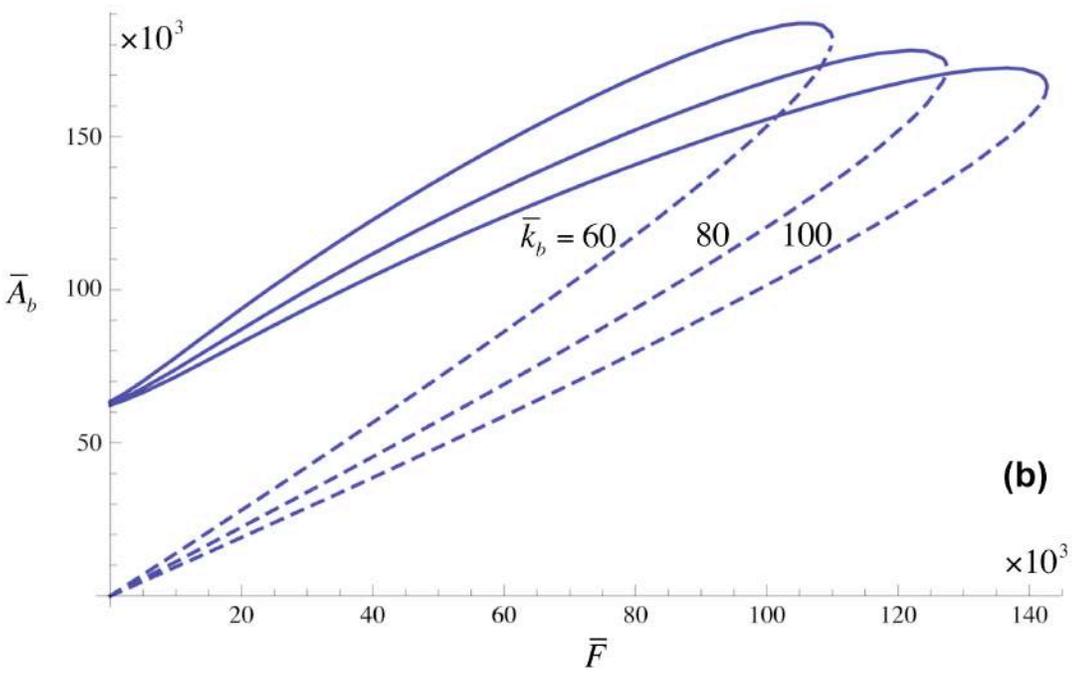

Figure 7

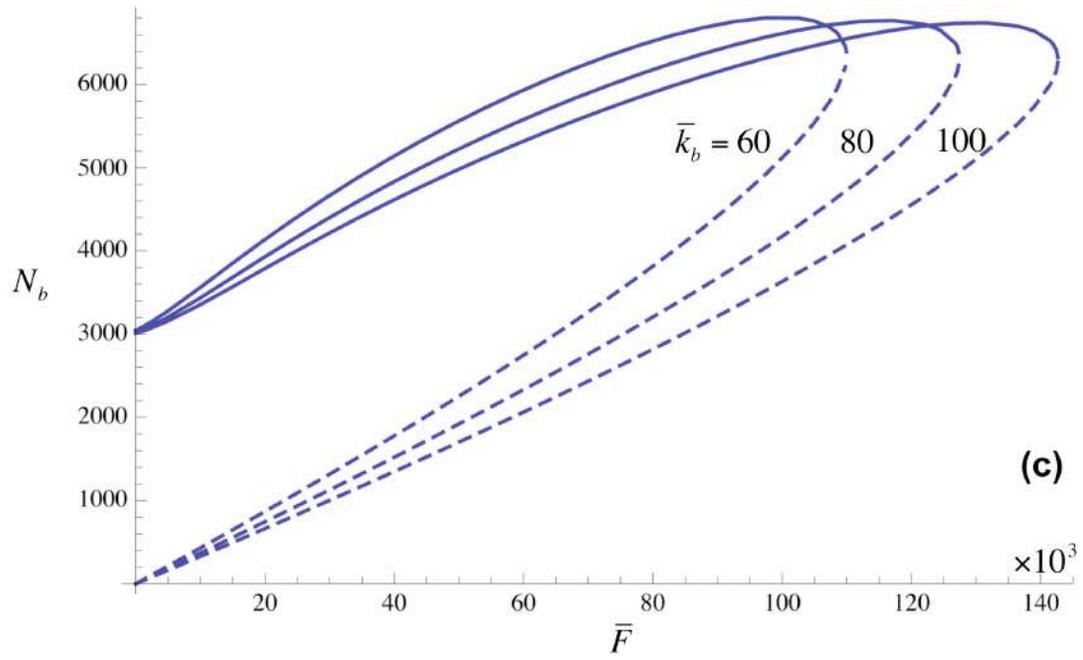

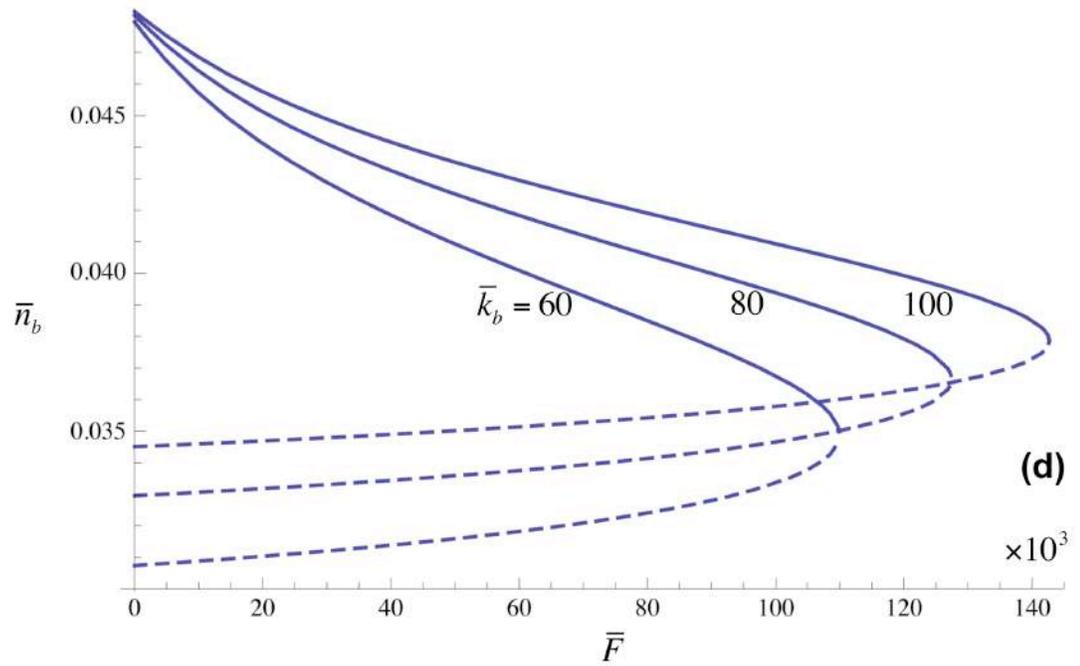

Figure 8

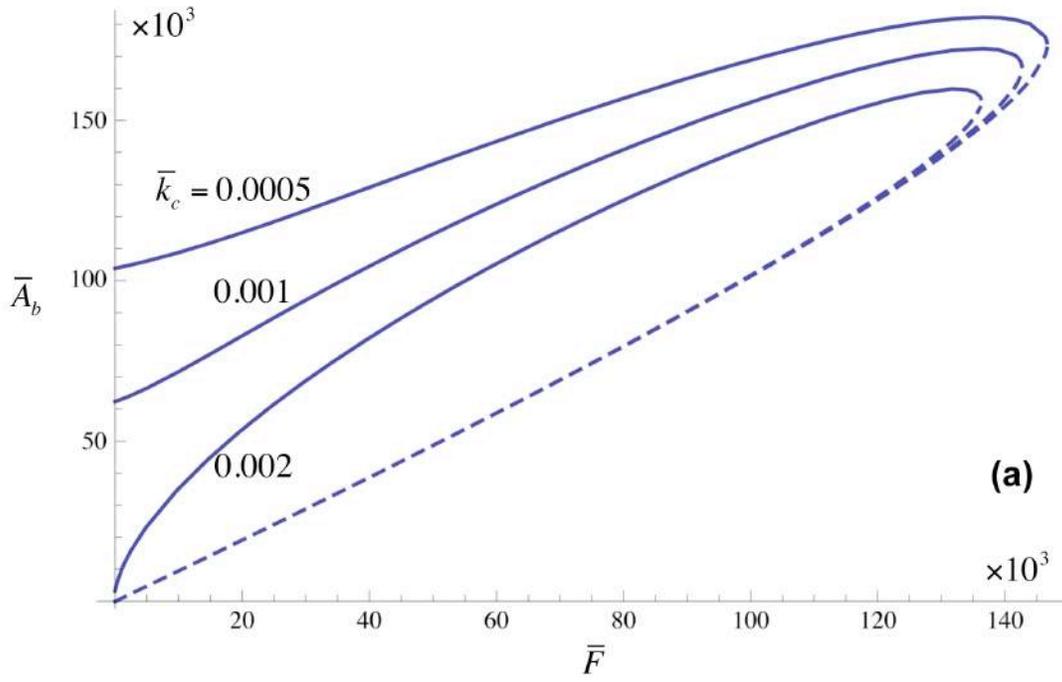

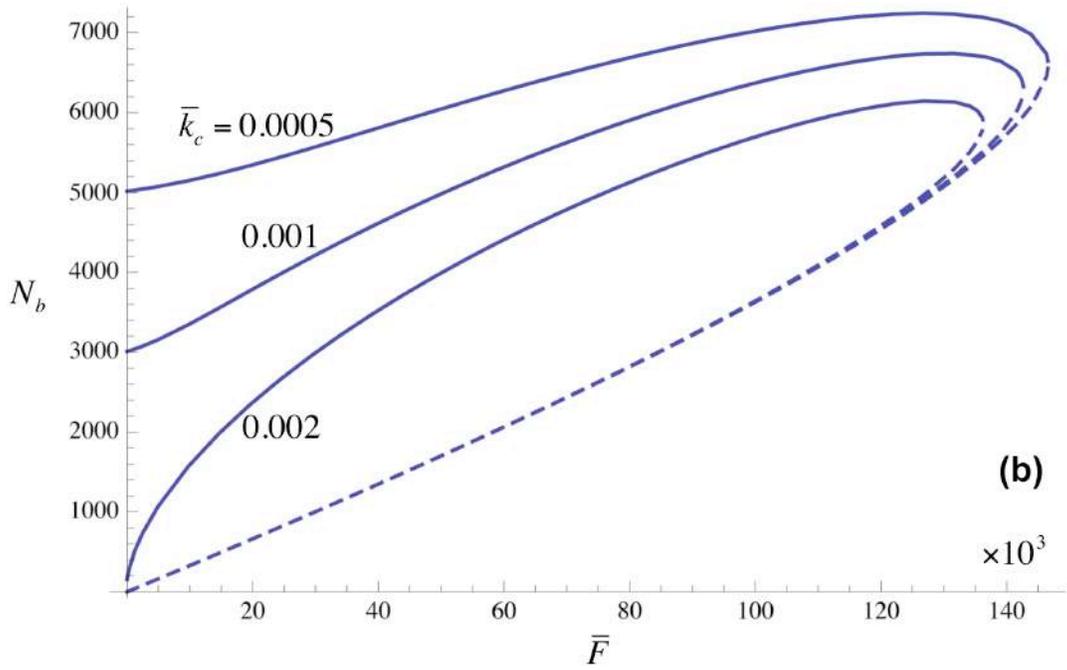

Figure 9

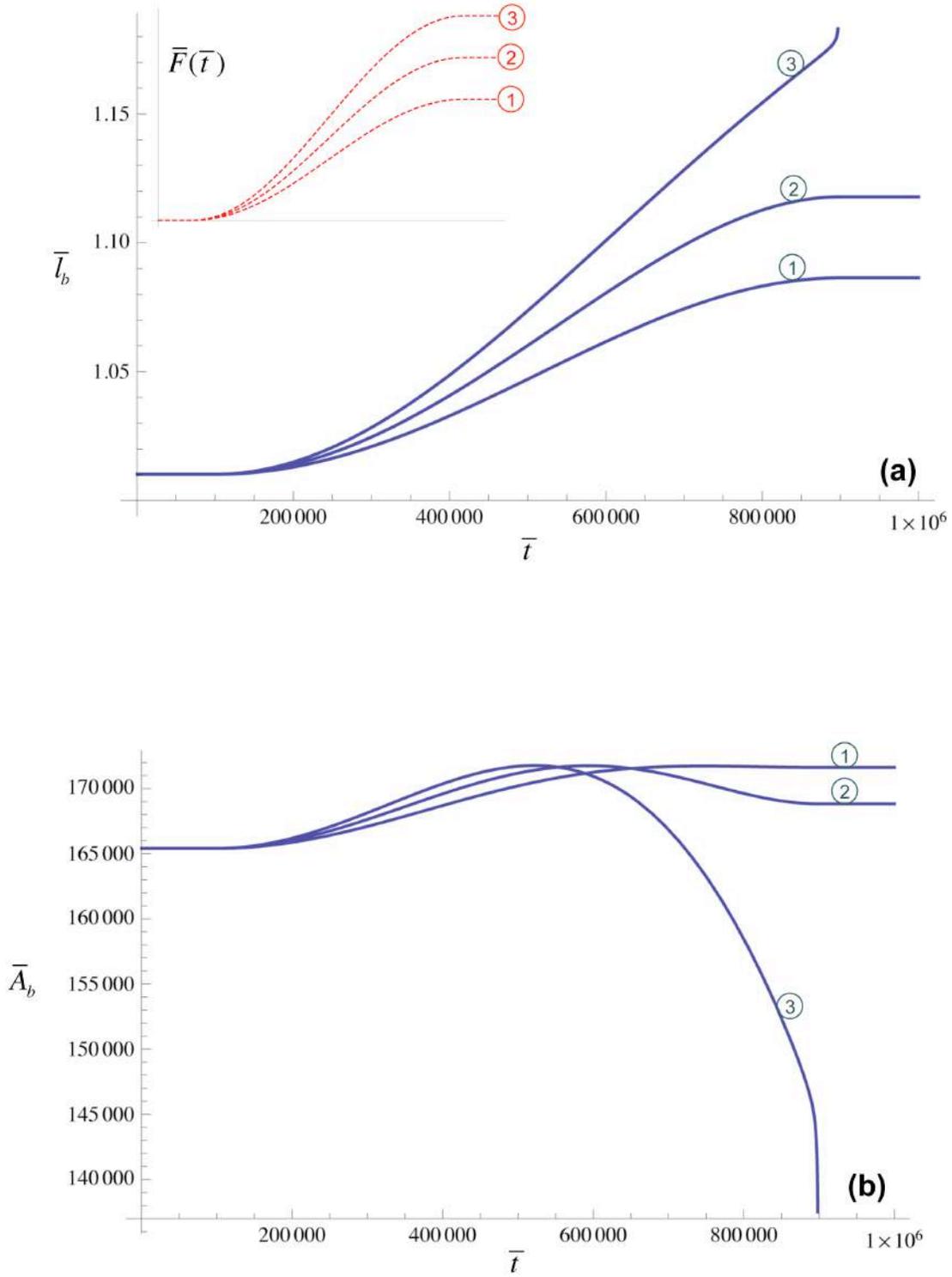

(a)

(b)

Figure 9

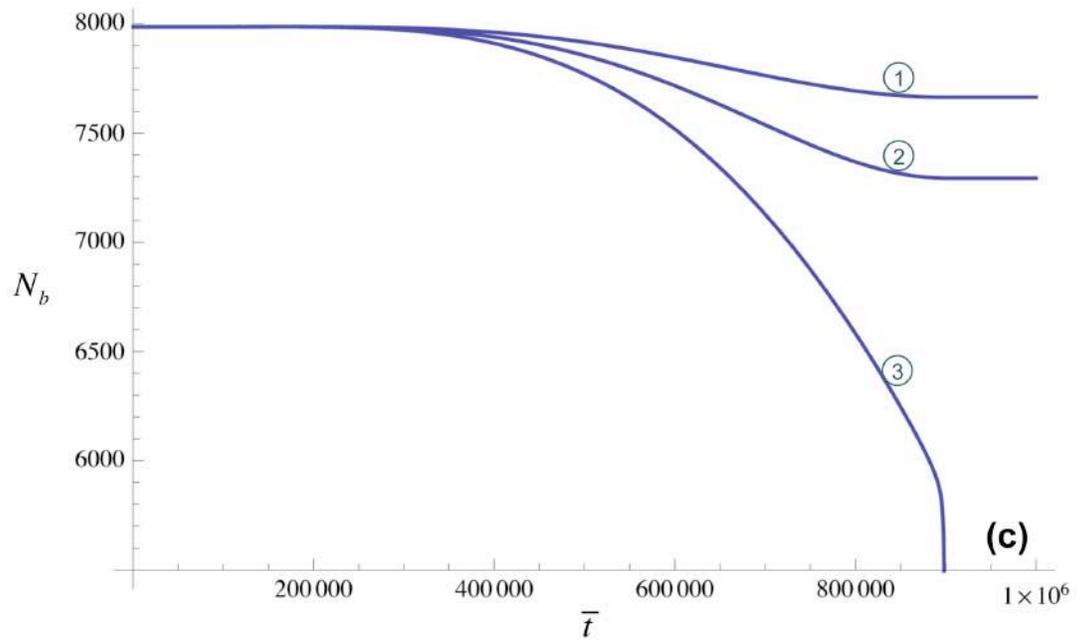

Figure 10

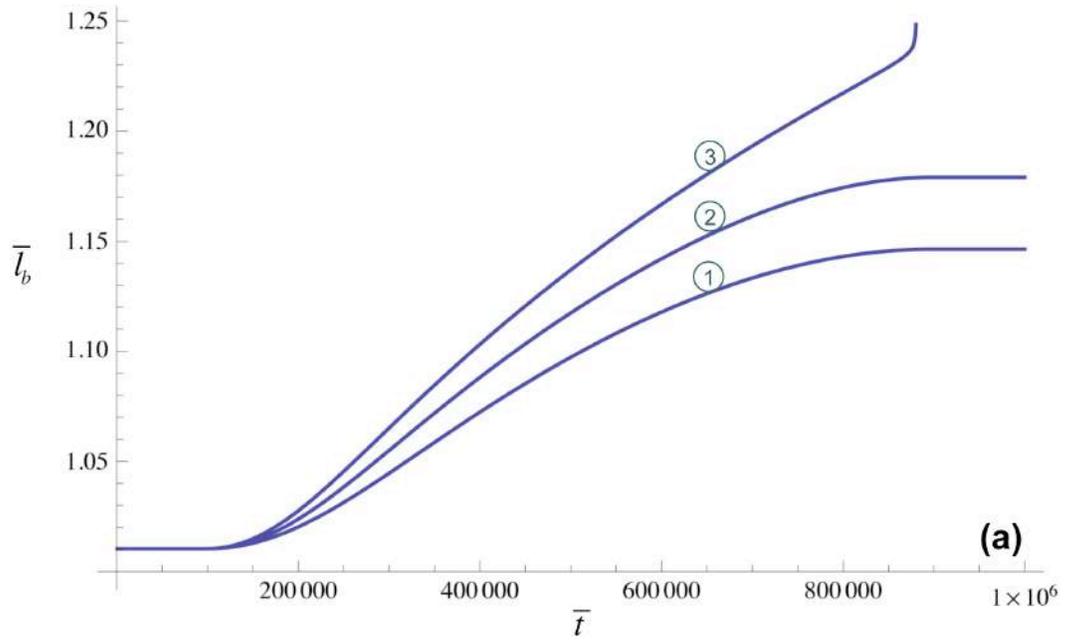

(a)

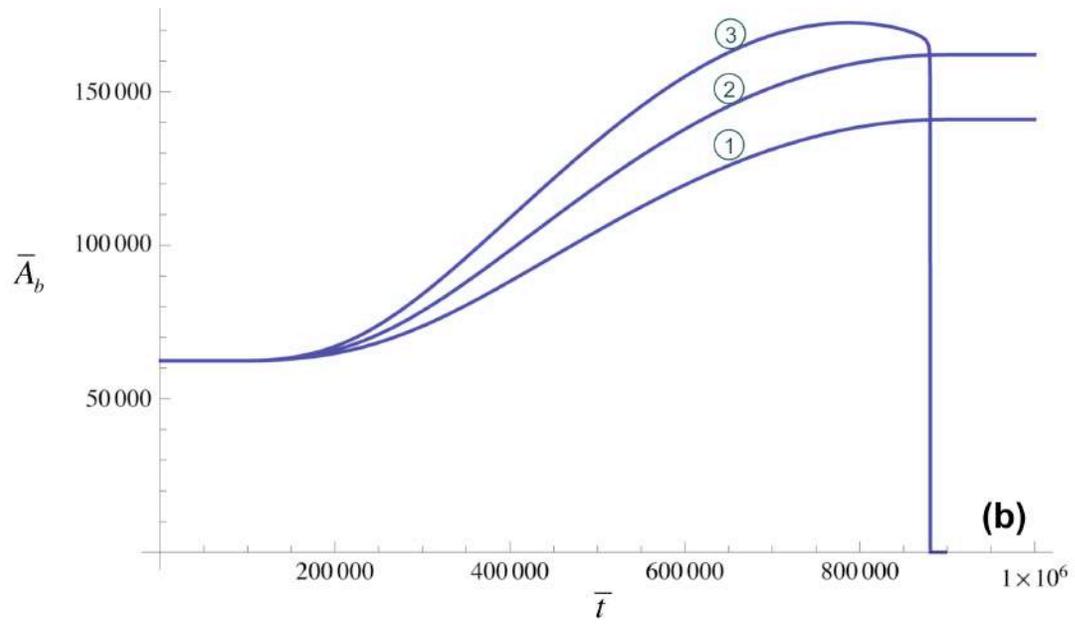

(b)

Figure 10

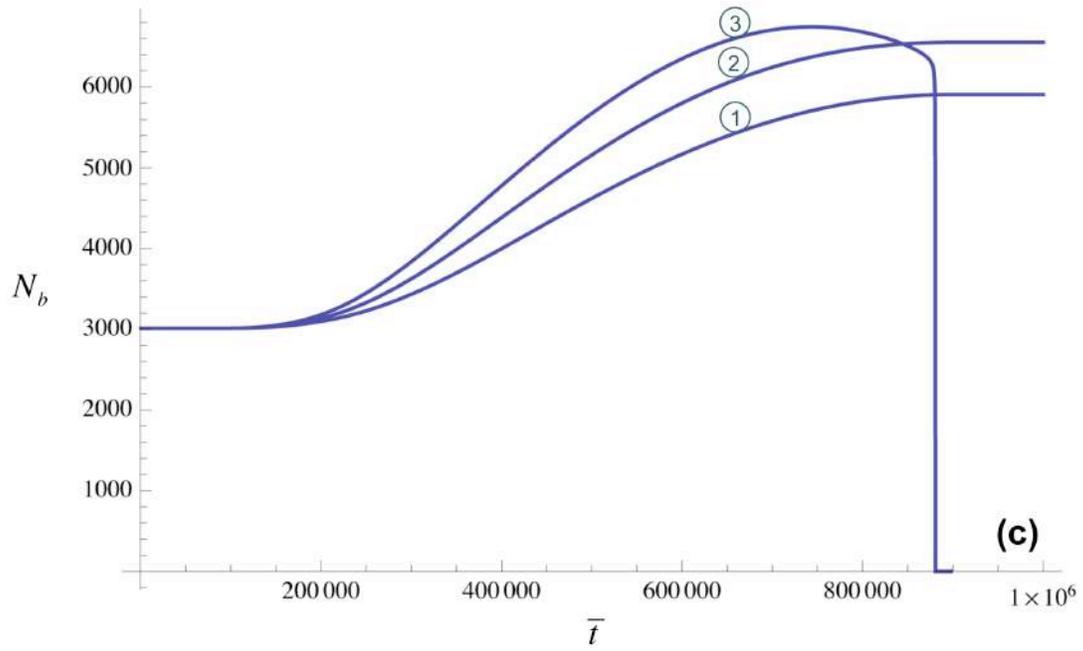

Figure 11

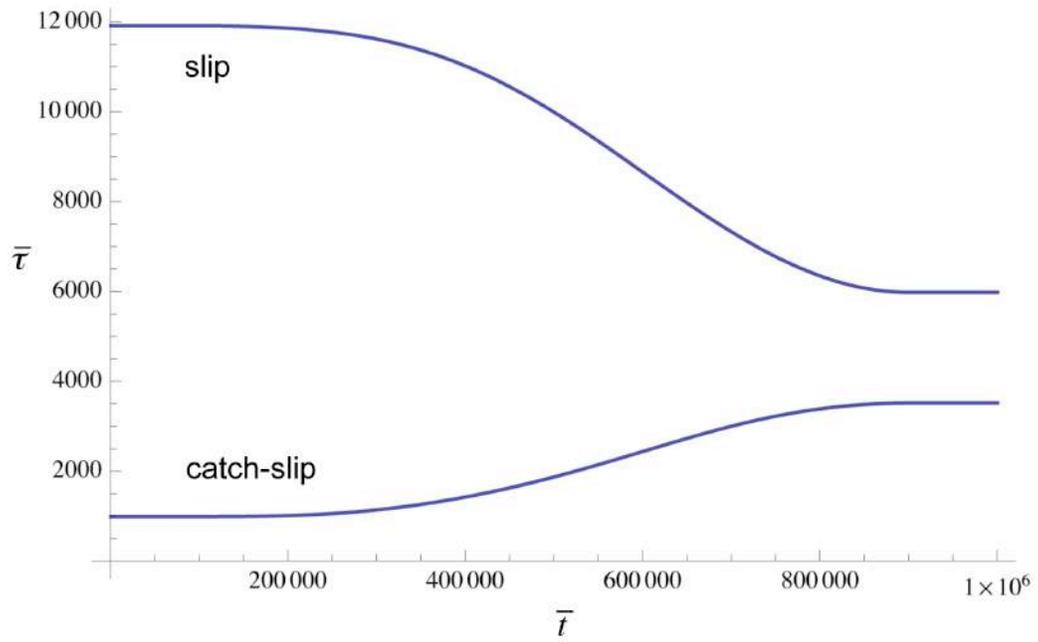

Figure 12

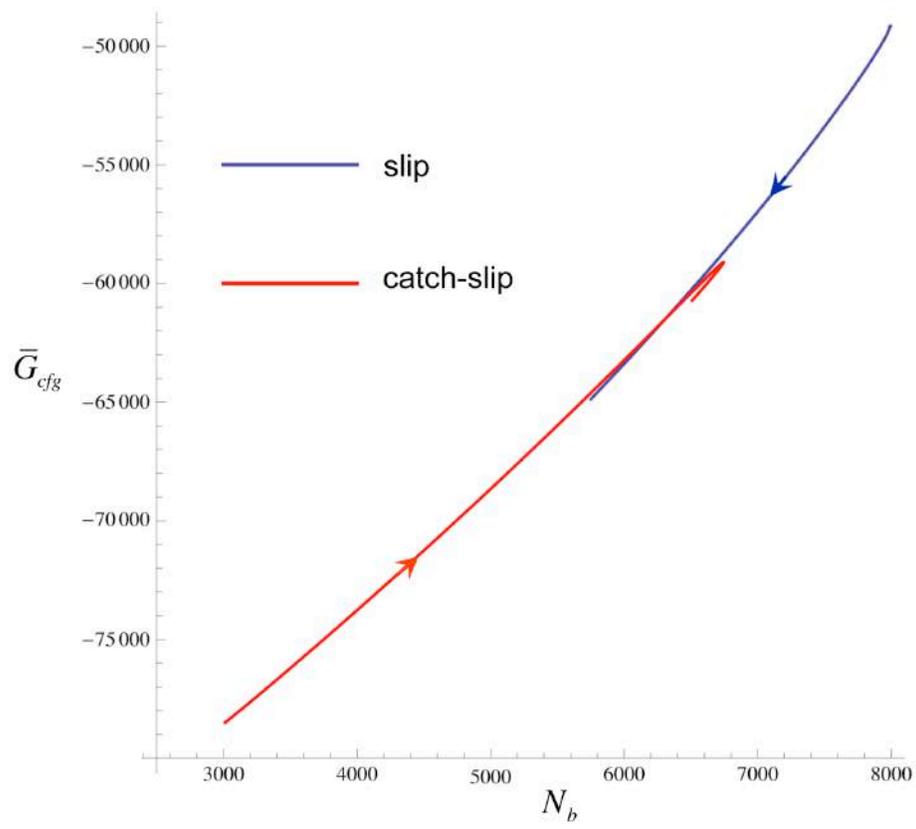